\documentclass[11pt]{article}

\usepackage{amsmath,amsfonts,amssymb,amsthm}
\usepackage{amsrefs}
\usepackage{xspace}
\usepackage{url} 

\newtheorem{theorem}{Theorem}
\newtheorem{lemma}[theorem]{Lemma}
\newtheorem{corollary}[theorem]{Corollary}

\newcommand{\bbC}{{\mathbb{C}}}
\newcommand{\C}{{\mathbb{C}}}

\newcommand{\bbZ}{{\mathbb{Z}}}

\newcommand{\cD}{{\mathcal{D}}}
\newcommand{\cN}{{\mathcal{N}}}
\newcommand{\cP}{{\mathcal{P}}}
\newcommand{\cQ}{{\mathcal{Q}}}
\newcommand{\cS}{{\mathcal{S}}}
\newcommand{\cU}{{\mathcal{U}}}
\newcommand{\cV}{{\mathcal{V}}}

\newcommand{\tr}{\mathop{\mathrm{tr}}}

\newcommand{\poly}{\mathop{\mathrm{poly}}}
\newcommand{\E}{\mathop{\mbox{$\mathbf{E}$}}}
\newcommand{\schur}{\mathop{\mathrm{Schur}}}
\newcommand{\schursmall}[2]{{\mathrm{S}}_{#1,#2}}
\newcommand{\planch}{\mathop{\mathrm{Planch}}}
\newcommand{\planchsmall}[1]{{\mathrm{P}}_{#1}}
\renewcommand{\d}{{\mathrm{d}}}

\newcommand{\HSP}{\textsc{hsp}\xspace}

\renewcommand{\>}{\rangle}
\newcommand{\<}{\langle}
\newcommand{\ket}[1]{|#1\rangle}
\newcommand{\bra}[1]{\langle #1|}
\newcommand{\scalar}[2]{\langle #1|#2\rangle}
\newcommand{\proj}[1]{\left|#1\right\>\!\left\<#1\right|}

\newcommand{\ot}{\otimes}
\newcommand{\eps}{\epsilon}
\newcommand{\ra}{\rightarrow}

\renewcommand{\setminus}{\mathbin{-}}
\newcommand{\partitionof}{\mathbin{\vdash}}

\newcommand{\be}{\begin{equation}}
\newcommand{\ee}{\end{equation}}
\def\ba#1\ea{\begin{align}#1\end{align}}
\newcommand{\nn}{\nonumber\\}
\newcommand{\eq}[1]{(\ref{eq:#1})}
\newcommand{\thmref}[1]{Theorem~\ref{thm:#1}}
\newcommand{\corref}[1]{Corollary~\ref{cor:#1}}
\newcommand{\lemref}[1]{Lemma~\ref{lem:#1}}
\newcommand{\secref}[1]{Section~\ref{sec:#1}}

\begin{document}


\title{\Large\textbf{
Weak Fourier-Schur sampling, \\
the hidden subgroup problem, \\
and the quantum collision problem}} 

\author{
Andrew M.\ Childs\footnote{amchilds@caltech.edu}\\[.5ex]
Institute for Quantum Information \\
California Institute of Technology \\
Pasadena, CA 91125, USA
\and
Aram W. Harrow\footnote{a.harrow@bris.ac.uk} \\[.5ex]
Department of Computer Science \\
University of Bristol \\
Bristol, BS8 1UB, UK
\and
Pawel Wocjan\footnote{wocjan@cs.ucf.edu} \\[.5ex]
School of Electrical Engineering and Computer Science \\
University of Central Florida \\
Orlando, FL 32816, USA
}

\date{}

\maketitle

\abstract{
Schur duality decomposes many copies of a quantum state into subspaces labeled by partitions, a decomposition with applications throughout quantum information theory.  Here we consider applying Schur duality to the problem of distinguishing coset states in the standard approach to the hidden subgroup problem.  We observe that simply measuring the partition (a procedure we call {\em weak Schur sampling}) provides very little information about the hidden subgroup.  Furthermore, we show that under quite general assumptions, even a combination of weak Fourier sampling and weak Schur sampling fails to identify the hidden subgroup.  We also prove tight bounds on how many coset states are required to solve the hidden subgroup problem by weak Schur sampling, and we relate this question to a quantum version of the collision problem.
}

\section{Introduction}
\label{sec:intro}

The hidden subgroup problem (\HSP) is a central challenge for quantum computation.  On the one hand, many of the known fast quantum algorithms are based on the efficient solution of the abelian \HSP \cites{Sho97,Hal02,Hal05,SV05}.  On the other hand, the {\em nonabelian} hidden subgroup problem has potential algorithmic applications: in particular, the graph isomorphism problem can be reduced to the \HSP in the symmetric group \cites{BL95,EH99}, and the shortest lattice vector problem can be reduced to the \HSP in the dihedral group (assuming it is solved by the standard method discussed below) \cite{Reg02}.  Unfortunately, no efficient algorithms are known for these two instances of the nonabelian hidden subgroup problem.  However, some partial progress has been made: there is a subexponential time algorithm for the dihedral \HSP \cites{Kup03,Reg04}, and it is known how to solve the \HSP efficiently for a variety of other nonabelian groups \cites{BCD05,FIMSS03,HRT00,Gav04,GSVV04,IMS03,MRRS04}.

In the hidden subgroup problem for a group $G$, we are given black-box access to a function $f:G \to S$, where $S$ is some finite set.  We say that $f$ hides a subgroup $H \le G$ provided $f(g)=f(g')$ if and only if $g^{-1} g' \in H$ (or in other words, provided $f$ is constant on left cosets of $H$ in $G$ and distinct on different left cosets).  The goal is to determine $H$ (say, in terms of a generating set) as quickly as possible.  In particular, we say that an algorithm for the \HSP in $G$ is efficient if it runs in time $\poly(\log |G|)$.

Nearly all quantum algorithms for the hidden subgroup problem make use of the so-called {\em standard method}, in which one queries the function $f$ on a uniform superposition of group elements and then discards the function value, giving a {\em coset state}
\be
  |gH\> := \frac{1}{\sqrt{|H|}} \sum_{h \in H} |gh\>
\ee
for some unknown, uniformly random $g \in G$.  In other words, the state is described by the density matrix
\ba
  \rho_H &:= \frac{1}{|G|} \sum_{g \in G} |gH\>\<gH| \\
         &=  \frac{1}{|G|} \sum_{h \in H} R(h)
\label{eq:hs_state}
\ea
(sometimes called a {\em hidden subgroup state}),
where $R$ is the {\em right regular representation} of $G$, satisfying
\be
  R(g) |g'\> = |g' g^{-1}\>
\ee
for all $g,g' \in G$.  Now the hidden subgroup problem is reduced to the problem of performing a measurement to distinguish the states $\rho_H$ for the various possible $H \le G$.

The symmetry of the hidden subgroup state can be exploited using
Fourier analysis.  In particular, the group algebra $\C G$ decomposes
under the left and right multiplication actions of $G$ (which commute,
and indeed generate each other's commutants) as
\be
  \C G \stackrel{G \times G}{\cong}
  \bigoplus_{\sigma \in \hat G} \cV_\sigma \otimes \cV_\sigma^*
\label{eq:fourier_duality}
\ee
where $\hat G$ denotes a complete set of irreducible representations (or {\em irreps}) of $G$, and $\cV_\sigma$ and $\cV_\sigma^*$ are the (row and column, respectively) subspaces acted on by $\sigma \in \hat G$.  The unitary transformation that relates the standard basis for $\C G$ and the basis for the spaces $\cV_\sigma \otimes \cV_\sigma^*$ is the Fourier transform, which can be carried out efficiently for most groups of interest \cites{Cop94,HH00,Bea97,MRR04}.

Since the hidden subgroup state $\rho_H$ is invariant under the left multiplication action of $G$, the decomposition \eq{fourier_duality} shows that it is block diagonal in the Fourier basis, with blocks labeled by the irreps $\sigma \in \hat G$.  For each $\sigma$, there is a $\dim \cV_\sigma \times \dim \cV_\sigma$ block that appears $\dim \cV_\sigma$ times (or in other words, the state is maximally mixed in the row space).  Thus, without loss of information (or in Kuperberg's words, ``sacrificing no entropy''), we can measure the irrep name $\sigma$ and discard the information about which $\sigma$-isotypic block occurred (sometimes referred to as ``discarding the row label'').

The process of measuring the irrep name $\sigma$ is referred to as {\em weak Fourier sampling}.  If $G$ is abelian, then all its irreps are $1$-dimensional.  In this case, weak Fourier sampling is a complete measurement, and all that remains is to use the statistics of the irrep name to infer the hidden subgroup.  However, for the nonabelian \HSP, some (indeed, typically most) irreps are more than $1$-dimensional, so weak Fourier sampling is an incomplete measurement.
Indeed, for most nonabelian groups, it turns out that weak Fourier sampling alone produces insufficient information to identify the hidden subgroup.  In particular, this is true of the \HSP relevant to graph isomorphism \cites{HRT00,GSVV04} as well as the dihedral \HSP.  To obtain further information about the hidden subgroup, one must perform a refined measurement inside the $(\dim \cV_\sigma)$-dimensional subspace that remains after weak Fourier sampling gives the result $\sigma$.  The process of making such a refined measurement is referred to as {\em strong Fourier sampling}.  There are many possible ways to measure inside $\cV_\sigma$, especially when it is high-dimensional, and much work has been done on the possible measurements for particular nonabelian groups.

Of course, with either weak or strong Fourier sampling, a single
hidden subgroup state is not sufficient to determine $H$: one must
repeat the sampling procedure to obtain statistics.  However, if one
simply repeats strong Fourier sampling a polynomial number of times,
there are some groups---in particular, the symmetric group---in which
hidden subgroups cannot be found, even if the measurement is allowed
to be chosen adaptively \cite{MRS05}.  Thus, to solve the \HSP in
general, one must perform a joint measurement on $k=\poly(\log |G|)$
copies of the hidden subgroup state, $\rho_H^{\otimes k}$.  In fact,
there are groups---again including the symmetric group---for which the
measurement must be entangled across $\Omega(\log |G|)$ copies
\cite{HMRRS06}.  Thus the difficulty of the general \HSP may be
attributed at least in part to that fact that highly entangled
measurements are required (although it should be noted that there are
some groups, such as the dihedral group, where single-register
measurements are at least information-theoretically sufficient
\cite{EH00}, yet an efficient algorithm remains elusive).  While it is
known that $O(\log |G|)$ copies are always information-theoretically
sufficient \cite{EHK99} (so that, in particular, the query complexity
of the \HSP is polynomial), there are many groups for which it is not
known how to  efficiently extract the identity of the hidden subgroup. 

Although previous work on the hidden subgroup problem has focused
almost exclusively on Fourier sampling, it turns out that there is
another kind of measurement that can also be performed without loss of
information.  The idea is to exploit the fact that the state
$\rho_H^{\otimes k}$ is symmetric under permutations of the $k$
registers.  Thus, we should consider the decomposition of the space
$(\C G)^{\otimes k}$ afforded by {\em Schur duality} \cite{GW98},
which decomposes $k$ copies of a $d$-dimensional space as 
\be
  (\C^d)^{\otimes k} \stackrel{\cS_k \times \cU_d}{\cong}
  \bigoplus_{\lambda \partitionof k}
  \cP_\lambda \otimes \cQ_\lambda^d
\label{eq:schur_duality}
\ee
where the symmetric group $\cS_k$ acts to permute the $k$ registers and the unitary group $\cU_d$ acts identically on each register.  Here the subspaces $\cP_\lambda$ and $\cQ_\lambda^d$ correspond to irreps of $\cS_k$ and $\cU_d$, respectively.  They are labeled by partitions $\lambda$ of $k$ (denoted $\lambda \partitionof k$), i.e., $\lambda=(\lambda_1,\lambda_2,\ldots)$ where $\lambda_1 \ge \lambda_2 \ge \ldots$ and $\sum_j \lambda_j = k$.
Note that we can restrict our attention to partitions with at most $d$ parts, since $\dim \cQ_\lambda^d = 0$ if $\lambda_{d+1}>0$.

Since $\rho_H^{\otimes k}$ is invariant under the action of $\cS_k$, the decomposition \eq{schur_duality} shows that it is block diagonal in the Schur basis with blocks labeled by $\lambda \partitionof k$.  For each $\lambda$, there is a $\dim \cQ_\lambda^{|G|} \times \dim \cQ_\lambda^{|G|}$ block that appears $\dim \cP_\lambda$ times (or in other words, the state is maximally mixed in the permutation space).  Thus, no information is lost if we measure the partition $\lambda$ and discard the permutation register.
By analogy to weak Fourier sampling, we refer to the process of
measuring $\lambda$ as {\em weak Schur sampling}.  This is a natural
measurement to consider not only because it can be performed without
loss of information, but also because it is a joint measurement of all $k$
registers, and we know that some measurement of this kind is required
to solve the general \HSP.  Unfortunately, we will see in
Section~\ref{sec:weak_schur} (and see also \corref{weak_schur} below) that
weak Schur sampling with $k=\poly(\log |G|)$ provides insufficient
information to solve the \HSP unless the hidden subgroup is very large
(in which case the problem is easy, even for a classical computer). 

In fact, since both weak Fourier sampling and weak Schur sampling can
be performed without loss of information, it is possible to perform
both measurements simultaneously (with the caveat that one must
discard the irrelevant information about the order in which the irreps
of $G$ appear).  Even though the statistics of the irrep name $\sigma$
and the partition $\lambda$ do not provide enough information to
identify the hidden subgroup, this does not preclude the possibility
that their joint distribution is more informative.  However, we will
see in Section~\ref{sec:weak_fourier_schur} that unless we are likely
to see the same representation more than once under weak Fourier
sampling (which is typically not the case), the Fourier and Schur
distributions are nearly uncorrelated.  Formally, we have
\begin{theorem}[Failure of weak Fourier-Schur sampling]
\label{thm:weak_fourier_schur}
The probability that weak Fourier-Schur sampling (defined in
\secref{weak_fourier_schur}) applied to $\rho_H^{\ot k}$ (defined in
\eq{hs_state}) 
provides a result that depends on $|H|$ is at most $k^2 |H| d_{\max}^2 /
|G|$, where $d_{\max}$ is the largest dimension of an irrep of $G$.
\end{theorem}
\noindent
This implies that $k$ needs to be large for most
cases of interest, including the dihedral and symmetric groups.
\begin{corollary}[Weak Fourier-Schur sampling on $\cD_N$ and $\cS_n$]
\label{cor:weak_fs_specific}
(a) Weak Fourier-Schur sampling on the dihedral group $\cD_N$ cannot
distinguish the trivial subgroup from a hidden reflection with
constant advantage unless $k = \Omega(\sqrt{N})$. (b) Weak
Fourier-Schur sampling on the symmetric group $\cS_n$ or on the
wreath product $\cS_n\wr\bbZ_2$ cannot
distinguish the trivial subgroup from an order $2$ subgroup with constant
advantage unless $k=\exp(\Omega(\sqrt{n}))$.
\end{corollary}

While it is unfortunate that weak Fourier-Schur sampling typically cannot provide enough information to find a hidden subgroup, it remains possible that Schur duality could be a helpful tool for solving the \HSP.  Just as weak Fourier sampling breaks the space into smaller subspaces in which one can perform strong Fourier sampling (or a multi-register generalization thereof), weak Fourier-Schur sampling provides an even finer decomposition into subspaces.  Since the latter decomposition produces subspaces that are entangled across all $k$ registers, it may be especially useful for {\HSP}s where we know that entangled measurements are necessary (or for which we would like to implement optimal measurements, which are typically entangled \cites{BCD05,BCD05b}).

The proof that weak Schur sampling fails is based on the simple
observation that distinguishing the trivial subgroup from a subgroup
of order $|H|$ in this way requires one to solve the problem of
distinguishing $1$-to-$1$ from $|H|$-to-$1$ functions on $G$, i.e.,
the $|H|$-collision problem for a list of size $|G|$.  Since there is
an $\Omega(\sqrt[3]{|G|/|H|})$ quantum lower bound for this problem
\cite{AS04}, $\poly(\log |G|)$ registers are insufficient.  In fact,
the problem resulting from the \HSP is potentially harder, since the
basis in which the collisions occur is inaccessible to the Schur
measurement.  This naturally leads to the notion of a {\em quantum}
collision problem, and raises the question of how quickly it can be
solved on a quantum computer.  In Section~\ref{sec:qcollision}, we define a sampling version of the quantum $r$-collision problem and use results on the asymptotics of the Plancherel measure on the symmetric group to prove that $k=\Theta(d/r)$ registers are necessary and sufficient to solve it.  In particular, we prove
\begin{theorem}[Quantum collision sampling problem]
\label{thm:nonadaptive}
Given $\rho^{\otimes k}$, distinguishing between [case $A$] $\rho =
I/d$ and [case $B$] $\rho^2 = \rho/\frac{d}{r}$
(i.e., $\rho$ is proportional to a projector of rank $d/r$) 
is possible with success probability $1 - \exp(-\Theta(kr/d))/2$.  In particular, success probability $\frac{1}{2}+\Omega(1)$ is possible if and only if $k=\Theta(d/r)$.
\end{theorem}
\noindent
In addition to providing the first results on estimation of the spectrum of a quantum state in the regime where $k \ll d^2$,
this gives tight estimates of the effectiveness of weak Schur
sampling, which we see requires an exponentially large (in $\log |G|$)
number of copies to be successful.
\begin{corollary}[Failure of Weak Schur sampling]
\label{cor:weak_schur}
  Applying Weak Schur sampling to $\rho_H^{\ot k}$ (where $\rho_H$ is
  defined in \eq{hs_state}) requires $k=\Theta(|G|/r)$ to distinguish
  the case $|H|\geq r$ from the case $H=\{1\}$ with constant
  advantage.
\end{corollary}
\noindent
The connection between \thmref{nonadaptive} and \corref{weak_schur} is
explained in \secref{weak_schur}, while \thmref{nonadaptive} is proved in
\secref{qcollision}.

Finally, in \secref{adaptive} we introduce a black box version of the quantum collision problem.  We show that it can be solved using $O(\sqrt[3]{d/r} \log d/r)$ queries, nearly matching the query lower bound from the classical problem (although the best algorithm we are aware of has a running time of $O(\sqrt{d/r} \log d/r)$).
We also discuss some alternative formulations of the quantum collision problem that are less closely connected to the \HSP, but that may be of independent interest.

\section{Weak Schur sampling}
\label{sec:weak_schur}

We begin by considering only the permutation symmetry of $k$ independent copies of the $d$-dimensional hidden subgroup states, without taking into account symmetry resulting from the group $G$.  In other words, we will consider only the Schur decomposition \eq{schur_duality}, and we will perform {\em weak Schur sampling}, i.e., a measurement of the partition $\lambda$.

The projector onto the subspace labeled by a particular $\lambda \partitionof k$ is
\be
  \Pi_\lambda := \frac{\dim \cP_\lambda}{k!} \sum_{\pi \in \cS_k}
                 \chi_\lambda(\pi) \, P(\pi)
\label{eq:Pi-lambda-def}
\ee
(see for example \cite{Ser77}*{Theorem 8}),
where  $\chi_\lambda$ is the character of the irrep of $\cS_k$ labeled
by $\lambda$ and $P$ is the (reducible) representation of $\cS_k$ that acts to permute the $k$ registers, i.e.,
\be
  P(\pi)|i_1\>|i_2\> \ldots |i_k\> =
  |i_{\pi^{-1}(1)}\>|i_{\pi^{-1}(2)}\>\ldots 
  |i_{\pi^{-1}(k)}\>
\ee
for all $i_1,\ldots,i_k\in\{1,\ldots,d\}$.  For any $d^k$-dimensional density matrix $\gamma$, the probability distribution under weak Schur sampling is
\be
  \Pr(\lambda|\gamma) =
  \tr(\Pi_\lambda \gamma)
\label{eq:schur_dist_gamma}
\,.
\ee

To use weak Schur sampling as part of a quantum algorithm, it is
important that the measurement of $\lambda$ can be carried out
efficiently.  The simplest implementation of the complete Schur
transform \cite{BCH06}, which fully resolves the subspaces $\cP_\lambda$
and $Q_\lambda^d$, runs in time $\poly(k,d)$, and thus is inefficient
when $d$ is exponentially large, as in the hidden subgroup problem.
It can be modified to run in time $\poly(k, \log d)$ either by a
relabeling trick \cite{Har05}*{footnote in Section 8.1.2} or by using
so-called {\em generalized phase estimation} \cites{BCH04, Har05}
(which may also be viewed as a generalization of the well-known swap
test \cites{BBDEJM96, BCWW01}).  Generalized phase estimation only
allows us to measure $\lambda$, but for weak Schur sampling this is
all we need.  In this procedure, we prepare an ancilla register in the
state $\frac{1}{\sqrt{k!}} \sum_{\pi \in \cS_k} |\pi\>$, use it to
perform a conditional permutation $P(\pi)$ on the input state
$\gamma$, and then perform an inverse Fourier transform over $\cS_k$
\cite{Bea97} on the ancilla register.  Measurement of the ancilla
register will then yield $\lambda \in \hat \cS_k$, interpreted as a
partition of $k$, distributed according to \eq{schur_dist_gamma}.

The distribution of $\lambda$ according to weak Schur sampling is invariant under the actions of the permutation and unitary groups, since these groups act only within the subspaces $\cP_\lambda$ and $\cQ_\lambda^d$, respectively.  In other words, for any $d^k$-dimensional density matrix $\gamma$, we have
\ba
  \Pr(\lambda|U^{\otimes k}\, \gamma\, U^{\dagger \, \otimes k}) 
  &= \Pr(\lambda|\gamma)
  \qquad \forall\, U \in \cU_d
\label{eq:unitary_invariance} \\
  \Pr(\lambda|P(\pi) \, \gamma \, P(\pi^{-1})) 
  &= \Pr(\lambda|\gamma)
  \qquad \forall\, \pi \in \cS_k
\label{eq:permutation_invariance}
\,.
\ea
(It is straightforward to verify these invariances by directly computing $[\Pi_\lambda,U^{\otimes k}]=[\Pi_\lambda,P(\pi)]=0$ and using the cyclic property of the trace.)  In particular, \eq{unitary_invariance} implies that for $\gamma = \rho^{\otimes k}$, the distribution according to weak Schur sampling depends only on the spectrum of $\rho$.  (In fact, it turns out that simply normalizing $\lambda$ provides a good estimate of the spectrum of $\rho$, provided $k$ is large \cite{KW01}.)

Now it is easy to see that weak Schur sampling on $k=\poly(\log |G|)$ copies of the hidden subgroup state $\rho_H$ provides insufficient information to solve the hidden subgroup problem.  The hidden subgroup state $\rho_H$ is proportional to a projector of rank $|G|/|H|$, as can be seen by computing
\be
  \rho_H^2 = \frac{1}{|G|^2} \sum_{h,h' \in H} R(h h')
           = \frac{|H|}{|G|} \rho_H
\,.
\ee
Since the distribution of measurement outcomes $\Pr(\lambda|\rho_H^{\otimes k})$ depends only on the spectrum of $\rho_H$, and the spectrum of $\rho_H$ depends only on the order of the hidden subgroup, it is clear that different subgroups of the same order cannot be distinguished by weak Schur sampling.  Nevertheless, we might hope to distinguish the trivial hidden subgroup from a hidden subgroup of order $|H| \ge 2$, which would be sufficient to solve the \HSP relevant to graph isomorphism, for example.  However, even this task requires an exponential number of hidden subgroup states, which can be seen as follows.

Suppose that weak Schur sampling could distinguish between hidden subgroup states corresponding to $H=\{1\}$ and some particular $H$ of order $|H| \ge 2$.  Since the distribution of $\lambda$ depends only on the spectrum of the state, this would mean that we could distinguish $k$ copies of the maximally mixed state $I_{|G|}/|G|$, where $I_d$ is the $d \times d$ identity matrix, from $k$ copies of the state $J_{|G|/|H|}/(|G|/|H|)$, where $J_{d'}$ is a projector onto an arbitrary subspace of dimension $d'$.  This in turn would imply that we could distinguish $1$-to-$1$ functions from $|H|$-to-$1$ functions using $k$ queries of the function.  Then the quantum lower bound for the $|H|$-collision problem \cite{AS04} shows that $k=\Omega(\sqrt[3]{|G|/|H|})$ copies are required.

Of course, this does not mean that $O(\sqrt[3]{|G|/|H|})$ copies are
sufficient.  In fact, it turns out that a linear number of copies is
both necessary and sufficient, as we will show by a more careful
analysis in Section~\ref{sec:qcollision}.  There we will prove
\thmref{nonadaptive}, which by the arguments of this section implies \corref{weak_schur}.

\section{Weak Fourier-Schur sampling}
\label{sec:weak_fourier_schur}

In the previous section, we showed that weak Schur sampling provides insufficient information to efficiently solve the hidden subgroup problem.  However, even though weak Fourier sampling typically also does not provide enough information, it is conceivable that the joint distribution of the two measurements could be substantially more informative.  In this section, we will see that this is not the case: provided weak Fourier sampling fails, so does weak Fourier-Schur sampling.

Because neither measurement constitutes a loss of information, it is
in principle possible to perform both weak Fourier sampling and weak
Schur sampling simultaneously.   If we perform weak Fourier
sampling in the usual way, measuring the irrep label for each
register, then we will typically obtain a state that is no longer
permutation invariant.  However, since the irrep labels are
identically distributed for each register, the order in which the
irreps appear carries no information.  Only the {\em type} of the
irreps, i.e., the number of times each irrep appears,
is important.  Thus, it is sufficient to perform what we might call
{\em weak Fourier type sampling}, in which we only measure the irrep
type.  Equivalently, we could perform complete weak Fourier sampling
and then randomly permute the $k$ registers, thereby discarding the
(irrelevant) information about the order of the irreps, and producing
a permutation-invariant state to use for weak Schur sampling.  (In
fact, since the distribution under weak Schur sampling has the
permutation invariance \eq{permutation_invariance}, it is not actually
necessary to explicitly symmetrize the state.) 

We begin by performing weak Fourier sampling.  The hidden subgroup state $\rho_H$ defined in \eq{hs_state} has the following block structure in the Fourier basis:
\ba
  \rho_H 
  &\cong  \frac{1}{|G|} \bigoplus_{\sigma\in\hat{G}}
           I_{\dim \cV_\sigma} \otimes
           \sum_{h\in H} \sigma(h)^* \\
  &=: \sum_{\sigma\in\hat G}
       \Pr(\sigma)\, 
       \frac{I_{\dim \cV_\sigma}}{\dim\cV_\sigma}\, \otimes\,
       \rho_{H,\sigma}
\,.
\ea
Here the probability of observing the irrep $\sigma$ is
\be
  \Pr(\sigma) = 
  \frac{\dim \cV_\sigma}{|G|}\sum_{h\in H}\chi_\sigma(h)^*
\label{eq:F-sampling-Pr-sigma}\,,
\ee
and the state conditioned on this observation is $\rho_{H,\sigma}$.
Weak Fourier sampling consists of measuring the label of the irrep.
Assume we have observed $\sigma$.  Then the resulting density operator
is 
\be
  \rho_{H,\sigma} =
  \frac{1}{\sum_{h\in H}
  \chi_\sigma(h)} 
  \sum_{h\in H}
  |\sigma\>\<\sigma| \otimes
  \sigma(h)^*
\,.
\ee
We explicitly include the label $\ket{\sigma}$ to reflect the fact that
$\sigma$ was observed.

Repeating weak Fourier sampling $k$ times, we obtain the state
\be
  \rho_{H,\underline{\sigma}} 
  = \rho_{H,\sigma_1}\otimes\cdots\otimes\rho_{H,\sigma_k}
\,,
\ee
where $\underline{\sigma}:=(\sigma_1,\sigma_2,\ldots,\sigma_k) \in \hat G^k$ may be viewed either as the actual outcome of the $k$ instances of weak Fourier sampling, or merely as a representative of the irrep type, as discussed above.  Given this state, the conditional probability of observing the partition $\lambda$ is
\ba
  \Pr(\lambda|\underline{\sigma})
  &= \tr(\Pi_\lambda\,\rho_{H,\underline{\sigma}}) \\
  &= \frac{\dim\cP_\lambda}{k!} 
     \sum_{\pi\in \cS_k}\chi_\lambda(\pi) 
     \tr[P(\pi)\,\rho_{H,\underline{\sigma}}]
\,.
\ea
Note that $\tr[P(\pi)\, \rho_{H,\underline{\sigma}}]=0$ if $\pi(\underline{\sigma})\neq\underline{\sigma}$, where $\pi(\underline{\sigma})=(\sigma_{\pi^{-1}(1)},\ldots,\sigma_{\pi^{-1}(k)})$.

Now we are ready to prove \thmref{weak_fourier_schur}:
\begin{proof}[Proof of \thmref{weak_fourier_schur}]
Let us assume that $\underline{\sigma}$ is multiplicity-free, i.e., that all the $\sigma_i$'s are different.  In this case the traces are zero for all $\pi \ne 1$, where $1$ is the identity element of $\cS_k$.  Then
\be
  \Pr(\lambda|\underline{\sigma})
  = \frac{\dim\cP_\lambda}{k!}\, \chi_\lambda(1)\, 
     \tr \rho_{H,\underline{\sigma}}
  = \frac{(\dim\cP_\lambda)^2}{k!}
\,,
\ee
which is nothing but the Plancherel distribution over $\hat \cS_k$, and which in particular is independent of the hidden subgroup $H$.  This shows that we cannot extract any information about $H$ provided that we have obtained a multiplicity-free $\underline{\sigma}$.

It remains to show that the probability of obtaining a type
$\underline{\sigma}$ that is not multiplicity-free is vanishingly
small.  This can be seen as follows.  From \eq{F-sampling-Pr-sigma},
we have 
\be
  \Pr(\sigma)\le\frac{(\dim\cV_\sigma)^2}{|G|}|H|
\ee
since the absolute value of the character $\chi_\sigma$ at any group element can be at most $\dim\cV_\sigma$.  If we perform weak Fourier sampling on two copies of $\rho_H$, then the probability of obtaining the same irrep twice is
\ba
  \Pr(\text{$2$ irreps same}) 
  = \sum_{\sigma \in \hat G} \Pr(\sigma)^2
  \le \max_{\sigma \in \hat G} \Pr(\sigma)
  \le \frac{d_{\max}^2}{|G|} |H|
\,,
\ea 
where $d_{\max} := \max_{\sigma \in \hat G} \dim \cV_\sigma$ denotes the maximal dimension of any irrep of $G$.
For $k$ copies, we can upper bound the probability that any two irreps out of the $k$ observed irreps are the same by a union bound.  We find 
\ba
  \Pr(\text{any $2$ irreps same}) 
  \le \binom{k}{2}\, \Pr(\text{$2$ irreps same})
  \le \binom{k}{2} \, \frac{d_{\max}^2}{|G|}\, {|H|}
\,.
\label{eq:multprob}
\ea
Therefore, provided $|H|$ and $d_{\max}$ are not too large, the
probability of obtaining the same irrep twice is small.  This
concludes the proof of \thmref{weak_fourier_schur}.
\end{proof}

We now focus on the important special cases of the dihedral and
symmetric groups to prove 
\corref{weak_fs_specific}.
\begin{proof}[Proof of \corref{weak_fs_specific}]
When $G$ is the dihedral group $\cD_N$ of order $2N$, we have
$d_{\max}=2$, so \eq{multprob} shows that weak Fourier-Schur sampling
cannot distinguish the trivial subgroup from a hidden reflection (a
subgroup of order $2$) with constant advantage unless
$k=\Omega(\sqrt{N})$.  Similarly, for the symmetric group $G=\cS_n$,
we have $d_{\max}\le \sqrt{n!}  \exp(-c\sqrt{n})$ for some constant
$c$ \cite{VK89}, which implies that weak Fourier-Schur sampling cannot
solve the \HSP relevant to graph isomorphism unless $k =
\exp(\Omega(\sqrt{n}))$.  Finally, we can express the irreps of the
wreath product $\cS_n\wr\bbZ_2$ in terms of the irreps of
$\cS_n$ \cite{Ser77}*{Section 8.2}, implying that $d_{\max}\le
2n!\exp(-2c\sqrt{n})$.  Since $|\cS_n\wr\bbZ_2| = 2(n!)^2$, we again
have exponentially small distinguishability unless $k =
\exp(\Omega(\sqrt{n}))$.
\end{proof}

In \cite{CW05} two of us considered an alternative quantum
approach to the graph isomorphism problem based on the nonabelian
hidden shift problem.  It can be shown that weak Fourier-Schur
sampling fails for similar reasons when applied to hidden shift
states instead of hidden subgroup states.  The hidden shift states
have a similar block structure, where the blocks correspond to
irreps of $G$ (although in the hidden shift case, the distribution of
$\sigma$ is exactly Plancherel over $\hat G$, so weak Fourier sampling
trivially fails).  Again, the probability of seeing the same irrep
twice is exponentially small, and unless some irrep appears twice, the
conditional distribution of $\lambda$ is equal to the Plancherel
distribution over $\hat \cS_k$.

\section{The quantum collision problem}
\label{sec:qcollision}

In Section~\ref{sec:weak_schur}, we saw that weak Schur sampling cannot efficiently solve the hidden subgroup problem since such a solution would imply an efficient solution of the collision problem.  In fact, the problem faced by weak Schur sampling is considerably more difficult, since no information is available about the basis in which the collisions occur.  This motivates quantum generalizations of the usual (i.e., classical) collision problem, which we study in this section.
 
The classical $r$-collision problem is the problem of determining
whether a black box function with $d$ inputs (such that $r$ divides
$d$) is either $1$-to-$1$ or
$r$-to-$1$.  This problem has classical (randomized) query complexity
$\Theta(\sqrt{d/r})$---as evidenced by the well-known birthday
problem---and quantum query complexity $\Theta(\sqrt[3]{d/r})$
\cites{BHT97,AS04}.  The classical algorithm is quite simple: after
querying the function on $O(\sqrt{d/r})$ randomly chosen inputs, there
is a reasonable probability that a collision will appear, provided one
exists.  The quantum algorithm is slightly more subtle, making use of
Grover's algorithm for unstructured search \cite{Gro96}.  In
particular, while the classical algorithm queries the black box
non-adaptively, it is essential for the quantum algorithm to make
adaptive queries. 

We begin in Section \ref{sec:nonadaptive} by considering what one might call the {\em quantum $r$-collision sampling problem}, which is simply the problem of deciding whether one has $k$ copies of the $d$-dimensional maximally mixed state or of a state that is maximally mixed on an unknown subspace of dimension $d/r$.  This is exactly the problem faced by the weak Schur sampling approach to the hidden subgroup problem, so our results on the quantum collision sampling problem allow us to put tight bounds on the effectiveness of weak Schur sampling for the \HSP.  It turns out that $k=\Theta(d/r)$ copies are necessary and sufficient to distinguish these two cases with constant advantage.
Interestingly, this measurement is entangled across all $k$ registers, and we do not know whether its performance can be matched by an adaptive sequence of single-register measurements, as is the case for asymptotically large $k$ \cite{BBMR05}.  However, we show that non-adaptively measuring $k$ sets of $O(1)$ registers each is much less effective, succeeding only if $k=\Theta(d^2/r^2)$.

A complete definition of the quantum collision problem requires us to specify a unitary black box that hides the function, and that allows us to make adaptive queries.  We propose one such definition in Section \ref{sec:adaptive}, and show that the resulting quantum $r$-collision problem can be solved in $O(\sqrt[3]{d/r}\log d/r)$ queries, nearly matching the $\Omega(\sqrt[3]{d/r})$ lower bound from the classical collision problem.  We also mention some possible alternative formulations of the problem and comment on their complexities.

\subsection{The quantum collision sampling problem}
\label{sec:nonadaptive}

This section is devoted to the proof of \thmref{nonadaptive}.  We consider the sampling version of the quantum collision problem, in which we are given $k$ samples of a quantum state that is either maximally mixed on a Hilbert space of dimension $d$, or is maximally mixed on an unknown (but fixed) subspace of dimension $d/r$.  We begin by showing the equivalence of this problem with some related problems.

\begin{lemma}\label{lem:nonadapt-equiv}
\begin{samepage}
The following tasks are equivalent:
\begin{enumerate}
\item[(a)] 
Distinguishing [case $A$] $(I/d)^{\ot k}$
          from [case $B$]
\be
  \int\!\d{U} \left( \sum_{j=1}^{d/r} U |j\>\<j| U^\dag \right)^{\otimes k}
\label{eq:dr-schur-state}
\ee
where $\d{U}$ is the Haar measure on the $d$-dimensional unitary group
$\cU_d$.
\item[(b)] 
Distinguishing [case $A$] $(I/d)^{\ot k}$
          from [case $B$] $\rho^{\ot k}$
for arbitrary $\rho$ satisfying $\rho^2=\rho/\frac{d}{r}$ (i.e., proportional to a projector of rank $d/r$).
\item[(c)]
Distinguishing [case $A$] $\rho_1^{\ot k}$
          from [case $B$] $\rho_2^{\ot k}$, where
$\rho_1,\rho_2$ are density matrices on $\C^{d'}$, $d'\geq d$, $\rho_1^2 = \rho_1/d$ and $\rho_2^2 = \rho_2/\frac{d}{r}$.
\end{enumerate}
\end{samepage}
Moreover, the optimal solution to each problem is to perform weak
Schur sampling (i.e., the measurement with operators
$\{\Pi_\lambda\}_{\lambda\partitionof k}$ defined in
\eq{Pi-lambda-def}) and then to perform classical processing on the
measurement outcome.  Both the quantum measurement and the classical
processing can be performed in time $\poly(k, \log d)$.
\end{lemma}

\begin{proof}
First note that (a) reduces to (b), which in turn reduces to (c). Thus equivalence of all three will follow from reducing (c) to (a). To do this, we will demonstrate an optimal strategy for (a) and show that it performs equally well when applied to (c).

Both states in (a) are invariant under the actions of $\cS_k$ and $\cU_d$, so they can each be written in the form $\sum_\lambda \Pr(\lambda|\gamma) \Pi_\lambda / \tr \Pi_\lambda$, where $\gamma$ is the state appearing either in case $A$ or case $B$, the $\Pi_\lambda$ are defined according to \eq{Pi-lambda-def}, and $\Pr(\lambda|\gamma)=\tr \Pi_\lambda \gamma$.  Thus weak Schur sampling (followed by classical processing of the outcome $\lambda$) is an optimal measurement strategy for problem (a).

Now observe from \eq{Pi-lambda-def} that $\Pi_\lambda$ can be defined as a weighted sum of permutations, and thus $\tr \Pi_\lambda \rho^{\otimes k}$ depends on $\rho$ only via the quantities $\tr\rho, \tr\rho^2, \ldots, \tr \rho^k$ that appear when evaluating $\tr P(\pi)\rho^{\ot k}$ for $\pi\in\cS_k$.  This means that $\Pr(\lambda|\gamma)$ is unchanged if we either embed $(\bbC^d)^{\ot k}$ into $(\bbC^{d'})^{\ot k}$ for $d'\geq d$, or if we replace $\rho^{\ot k}$ with $(U\rho U^\dag)^{\ot k}$, or indeed with $\E_U(U\rho U^\dag)^{\ot k}$ for any distribution over unitaries $U$.  This implies that $\Pr(\lambda|\gamma)$ is the same for tasks (a), (b) and (c), and that any solution for (a) implies an equally effective solution for (c).

Finally, the measurement $\{\Pi_\lambda\}_{\lambda\partitionof k}$ (i.e., weak Schur sampling) can be efficiently implemented using generalized phase estimation, as described in \secref{weak_schur}.  The classical processing consists of comparing the conditional probabilities $\tr \Pi_\lambda \rho^{\ot k}$, for which we give efficiently calculable formulas below.
\end{proof}

Now we turn to the calculation of $\Pr(\lambda) = \tr\Pi_\lambda
\rho^{\ot k}$ for $\rho$ satisfying $\rho^2 = \rho / d$ (and embedded
into an ambient space of arbitrary dimension $d'\geq d$).   We call the
resulting distribution of $\lambda \partitionof k$ the {\em Schur distribution}, $\schur(k,d)$, in which $\lambda$ appears with probability
\ba
  \Pr(\lambda) &= \frac{\dim \cP_\lambda \dim \cQ_\lambda^d}{d^k} \\
               &= \frac{(\dim \cP_\lambda)^2}{k!}
                  \prod_{(i,j) \in \lambda} \! \left(1 + \frac{j-i}{d}\right)
\,,
\label{eq:schur-dist}
\ea
where in the second line we have used Stanley's formula for $\dim
\cQ_\lambda^d$  \cite{Sta71}, interpreting $\lambda$ as a Young diagram,
where $(i,j) \in \lambda$ iff $1 \le j \le \lambda_i$.

Our goal is to show that the distributions $\schur(k,d)$ and
$\schur(k,d/r)$ are close unless $k$ is $\Omega(d/r)$.  We will do
this by comparing the Schur distribution to the Plancherel
distribution over $\lambda \in \hat \cS_k$, $\planch(k)$, for which 
\be
  \Pr(\lambda) = \frac{(\dim \cP_\lambda)^2}{k!}
\,.\label{eq:planch-dist}
\ee
Note that we can think of $\planch(k)$ as the $d \to \infty$ limit of
$\schur(k,d)$.

The key technical part of the proof of \thmref{nonadaptive} is
the following lemma:

\begin{lemma}\label{lem:schur-planch}
If $2\leq k\leq d$ then
\be \frac{k}{36d} \leq 
\|\schur(k,d)-\planch(k)\|_1 \leq
  \sqrt{2}\frac{k}{d}
\,.
\ee
\end{lemma}

We note in passing that the upper bound can be amplified when $k\geq d$ as follows:
\begin{corollary}\label{cor:schur-planch-large-k}
\be
  \frac{1}{2}\|\schur(k,d)-\planch(k)\|_1 \geq 1 - e^{-\frac{1}{10368}
  \left(\frac{k}{d}-1\right)}
\,.
\ee
\end{corollary}

\begin{proof}[Proof of \corref{schur-planch-large-k}]
For density matrices $\rho, \sigma$, we define the fidelity
$F(\rho,\sigma):=\|\sqrt{\rho}\sqrt{\sigma}\|_1^2$ and
recall the relations \cite{FG97}
\be
  1-\sqrt{F(\rho,\sigma)} \leq
  \frac{1}{2}\|\rho-\sigma\|_1 \leq \sqrt{1-F(\rho,\sigma)}
\,,
\label{eq:fidelity-trace-dist}
\ee
where the second inequality is saturated when $\rho$ and $\sigma$ are
pure states.
Now use \lemref{schur-planch} to obtain
\be
  \sqrt{1-F(\schur(d,d),\planch(d))}
  \geq \frac{1}{2}\|\schur(d,d)-\planch(d)\|_1 \geq \frac{1}{72}
\,.
\label{eq:schur-planch-k-eq-d}
\ee
Next,
\ba
  F(\schur(k,d), \planch(k))
  &\leq F(\schur(d,d), \planch(d))^{\lfloor\frac{k}{d}\rfloor} \\
  &\leq \left(1-\frac{1}{5184}\right)^{\lfloor\frac{k}{d}\rfloor}
   \leq e^{-\frac{1}{5184}\left(\frac{k}{d}-1\right)}
\,,
\ea
using first that fidelity is nondecreasing under quantum operations
and is multiplicative under tensor products, then
\eq{schur-planch-k-eq-d}, and then
the inequality $1-x \leq e^{-x}$.
Finally,
\be
  \frac{1}{2}\|\schur(k,d)-\planch(k)\|_1
  \geq 1 - \sqrt{F(\schur(k,d), \planch(k))}
  \geq 1 - e^{-\frac{1}{10368}\left(\frac{k}{d}-1\right)}
\,,
\ee
which completes the proof.
\end{proof} 

To derive \thmref{nonadaptive} from
\lemref{schur-planch}, we need one additional fact.
\begin{lemma}[Monotonicity]\label{lem:monotonicity}
For any integers $k, d_1, d_2\geq 1$ and $r\geq 2$,
\be
  \left\|\schur(k,d_1) - \schur(k,d_2)\right \|_1  \geq
  \left\|\schur(k,rd_1) - \schur(k,rd_2)\right \|_1
\,.
\label{eq:monotonicity}
\ee
\end{lemma}

\begin{proof}
Define $\sigma_d$ (for $d$ one of $d_1, d_2, rd_1, rd_2$) as the state
\be
  \sigma_d := \int\!\d{U} \left( \sum_{j=1}^{d} U |j\>\<j| U^\dag \right)^{\otimes k}
\ee
where $\d{U}$ is the Haar measure on the $d'$-dimensional unitary
group $\cU_{d'}$ and $d'$ is some arbitrary integer satisfying $d'\geq
r\cdot\max(d_1, d_2)$.  Let $\cN$ be the quantum operation which takes
as input a state on $(\bbC^{d'})^{\ot k}$, adjoins the maximally mixed
state $(I_r/r)^{\otimes k}$, draws a uniform $U\in\cU_{rd'}$, applies $U^{\ot k}$, and finally outputs a state on
$(\bbC^{rd'})^{\ot k}$. Adjoining $I_r/r$ will map a state
proportional to a rank $d$ projector to a state proportional to a rank
$rd$ projector.  Thus, $\cN(\sigma_d) = \sigma_{rd}$ (up to a change
in the ambient dimension, which is irrelevant by
\lemref{nonadapt-equiv}). Then 
\eq{monotonicity} follows from the non-increase of trace distance
under quantum operations. 
\end{proof}
\noindent
(It seems plausible that \lemref{monotonicity} should also hold for
non-integer $r$ whenever $rd_1$ and $rd_2$ are integers, but we know
of no proof of this claim.)

Now we turn to the proof of \thmref{nonadaptive}.

\begin{proof}[Proof of \thmref{nonadaptive}]
The case when $k\leq d$ is the simple one.  Using first the triangle inequality, and then \lemref{schur-planch}, we find
\ba
  &\left\|\schur(k,d) - \schur(k,d/r)\right\|_1 \nn
  &\quad\le
  \left\|\schur(k,d) - \planch(k)\right\|_1 +
  \left\|\schur(k,d/r) - \planch(k)\right\|_1\\ 
  &\quad\leq \sqrt{2}\frac{k + kr}{d} \leq O(kr/d)
\,.
\ea
This establishes the $\Omega(d/r)$ lower bound on the number of copies.

Now, for any choice of $d$ and $r$, we would like to show that $k=d/r$ copies is sufficient to obtain $\|\schur(k,d) - \schur(k,d/r)\|_1 \geq \Omega(1)$.  The achievability part of \thmref{nonadaptive} will then follow from the same sort of amplification argument we used in the proof of \corref{schur-planch-large-k}. Choose $\nu$ to be the smallest positive integer satisfying $r^\nu \geq 72\sqrt{2}$; note that $\nu \leq 7$.  Then we use both the upper and lower bounds of \lemref{schur-planch} to find
\ba
\left\|\schur\left(\frac{d}{r}, \frac{d}{r}\right)-
\planch\left(\frac{d}{r}\right)\right\|_1 
& \geq \frac{1}{36} \\
\left\|\schur\left(\frac{d}{r}, dr^{\nu-1}\right) - 
\planch\left(\frac{d}{r}\right)\right\|_1 
 & \leq \frac{\sqrt{2}}{r^\nu} \leq \frac{1}{72}
\,.
\ea
Next we abbreviate $\schursmall{k}{d}:=\schur(k,d)$ and
$\planchsmall{k}:=\planch(k)$ and use the triangle inequality to bound
\ba
&\left\|\schursmall{\frac{d}{r}}{\frac{d}{r}}-
\schursmall{\frac{d}{r}}{d}\right\|_1  +
\left\|\schursmall{\frac{d}{r}}{d}-
\schursmall{\frac{d}{r}}{dr}\right\|_1  +
\cdots +
\left\|\schursmall{\frac{d}{r}}{dr^{\nu-2}}-
\schursmall{\frac{d}{r}}{dr^{\nu-1}}\right\|_1 
\label{eq:nonincreasing-terms}\\
&\quad\geq
\left\|\schursmall{\frac{d}{r}}{\frac{d}{r}} -
\planchsmall{\frac{d}{r}}\right\|_1  -
\left\|\schursmall{\frac{d}{r}}{dr^{\nu-1}} - 
\planchsmall{\frac{d}{r}}\right\|_1 
\geq \frac{1}{72}
\,.\nonumber
\ea
According to \lemref{monotonicity}, the first term on the left hand
side of \eq{nonincreasing-terms} is the largest.  Thus,
\be
\left\|\schur\left(\frac{d}{r}, \frac{d}{r}\right)-
\schur\left(\frac{d}{r}, d\right)\right\|_1 \geq \frac{1}{72\nu}
\geq \frac{1}{504}
\,.
\ee
This concludes the proof of \thmref{nonadaptive}.
\end{proof}

Our only remaining task is to prove 
\lemref{schur-planch}, establishing tight bounds on the distance
between $\schur(k,d)$ and $\planch(k)$.  Here our primary tools are
explicit formulas for the Plancherel averages of certain
functions of Young diagrams over $\cS_k$.

\begin{proof}[Proof of \lemref{schur-planch}]
Let $\Delta_{k,d}$
denote the total variation distance between 
$\schur(k,d)$ and $\planch(k)$.   According to \eq{schur-dist} and
\eq{planch-dist}, $\Delta_{k,d}:=\|\schur(k,d)-\planch(k)\|_1$ can be written 
\be
  \Delta_{k,d} 
  = \E_{\lambda \partitionof k} \left|
    \prod_{(i,j) \in \lambda} \left(1+\frac{j-i}{d}\right) - 1
    \right|
\label{eq:Delta-kd}\ee
where the expectation is taken over $\planch(k)$.

We will now derive both upper and lower bounds on $\Delta_{k,d}$ from
\eq{Delta-kd}, starting with the $O(k/d)$ upper bound.
By the Cauchy-Schwarz inequality (or concavity of the square root and Jensen's inequality),
\ba
  \Delta_{k,d}^2
  &\le \E_{\lambda \partitionof k} \left[
       \prod_{(i,j) \in \lambda} \left(1+\frac{j-i}{d}\right) - 1
       \right]^2 \\
  &=   \E_{\lambda \partitionof k} \left[
       \prod_{(i,j) \in \lambda} \left(1+\frac{j-i}{d}\right)^2 
       -2 \prod_{(i,j) \in \lambda} \left(1+\frac{j-i}{d}\right) + 1
       \right] \\
  &=   \E_{\lambda \partitionof k}
       \prod_{(i,j) \in \lambda} \left(1+\frac{j-i}{d}\right)^2 - 1
\,,
\ea
where in the last line we have used linearity of expectation and the fact that $\schur(k,d)$ is normalized.
Now using the inequality $1+x \le e^x$, we have
\ba
  \Delta_{k,d}^2
  &\le \E_{\lambda \partitionof k} \exp\!\left(
       2\sum_{(i,j) \in \lambda} \frac{j-i}{d} 
       \right) - 1 \\
  &=   \sum_{m=1}^{\infty} \frac{2^m}{m! \, d^m}
       \E_{\lambda \partitionof k} v_1(\lambda)^m
\label{eq:taylorbound}
\ea
where
\be
  v_1(\lambda) := \sum_{(i,j) \in \lambda} (j-i)
\ee
can be viewed as a measure of the asymmetry of $\lambda$.

Intuitively, we expect this to be small since typical Plancherel-distributed Young diagrams are close to a certain limiting shape that is symmetric, and in which the length of the longest row, $\lambda_1$, is close to $2 \sqrt k$ \cites{LS77,VK77}  (in fact, $\lambda_1$ is exponentially tightly concentrated around this value \cite{BDJ99}).  Our proof will use results obtained by Kerov in the course of describing the asymptotically Gaussian fluctuations about the limiting shape \cite{Ker93}.

It turns out that $v_1$ can be expressed exactly in terms of a central character of the symmetric group \cite{Juc74}, namely
\be
  v_1(\lambda) = \binom{k}{2} \frac{\chi_\lambda(\tau)}{\dim \cP_\lambda}
\ee
where $\tau$ is any representative of the conjugacy class of transpositions.
Then using the results of Kerov \cite{Ker93}*{equation (2.5) and Lemma 2.3}, we find
\ba
  \E_{\lambda \partitionof k} v_1(\lambda)^{2m-1} &= 0 \\
  \E_{\lambda \partitionof k} v_1(\lambda)^{2m}
  &= \frac{(2m)! \, k!}{4^m \, m! \, (k-2m)!}
     \le \frac{(2m)! \, k^{2m}}{4^m \, m!}
\label{eq:keroveven}
\ea
for $m=1,2,\ldots$.
Using these expressions in (\ref{eq:taylorbound}), we find
\ba
  \Delta_{k,d}^2
   \le \sum_{m=1}^{\infty} \frac{1}{m!} \left(\frac{k}{d}\right)^{2m}
   =   e^{(k/d)^2} - 1 
   \le 2 \left(\frac{k}{d}\right)^2
\,,
\ea
where in the second inequality we have assumed $k\leq d$.  This shows that
$\Delta_{k,d}=o(1)$ provided $k=o(d)$, and indeed, that $\Delta_{k,d}$
is exponentially small (in $\log d$) if $k = d^{1-\epsilon}$ for any
fixed $\epsilon>0$. 

Now we establish the $\Omega(k/d)$ lower bound for $\|\planch(k) -
\schur(k,d) \|_1$, again by a careful examination of the distribution
of $v_1(\lambda)$.
If $S$ is any subset of the partitions of $k$, then
\ba
  \Delta_{k,d}
  &\ge \E_{\lambda \partitionof k}\left[1-
       \prod_{(i,j) \in \lambda}\left(1+\frac{j-i}{d}\right)\right]
       \delta(\lambda \in S)
       \\
  &=   \Pr(\lambda \in S) - \E_{\lambda \partitionof k}
       \prod_{(i,j) \in \lambda}\left(1+\frac{j-i}{d}\right)
       \delta(\lambda \in S) \\
  &\ge \Pr(\lambda \in S) - \E_{\lambda \partitionof k}
       e^{v_1(\lambda)/d} \, \delta(\lambda \in S) \\
  &\ge \Pr(\lambda \in S) 
       \left(1 - \max_{\lambda \in S} e^{v_1(\lambda)/d}\right)
\,.
\ea
Thus, if we can show that $v_1(\lambda)$ has a reasonable probability
of being very negative, we will find that $\Delta_{k,d}$ is
appreciable.  Under the Plancherel distribution, the probability of a
diagram $\lambda$ and its transposed diagram $\lambda^T$ (satisfying
$v_1(\lambda^T) = - v_1(\lambda)$) are equal, so by symmetry,
\ba
  \Pr(v_1(\lambda) \le -v) &= \frac{1}{2} \Pr(v_1(\lambda)^2 \ge v^2)
\,. \ea
Letting $S$ be the set of $\lambda$'s with $v_1(\lambda) \le -v$ (for
some $v>0$ to be specified later), we have
\be
  \Delta_{k,d} \ge \frac{1}{2} \Pr(v_1(\lambda)^2 \ge v^2) (1-e^{-v/d})
\,.
\label{eq:v1bound}
\ee

Now we use \eq{keroveven} for $m=1$ and $m=2$, i.e.,
\ba
  \E_{\lambda \partitionof k} v_1(\lambda)^2 &= \binom{k}{2} \\
  \E_{\lambda \partitionof k} v_1(\lambda)^4 &= \frac{3}{4}k(k-1)(k-2)(k-3)
\,,
\ea
to show that
\be
\E_{\lambda \partitionof k} v_1(\lambda)^4 - 
\left[\E_{\lambda \partitionof k} v_1(\lambda)^2\right]^2 =
\frac{1}{4} k(k-1)(2k^2 - 15k + 18) \leq \frac{k^4}{2}
\,.
\ee
Applying a one-sided variant of the Chebyshev inequality, namely \cite{Sel40}
\be
  \Pr(X \ge \mu - \alpha \sigma) \ge \frac{\alpha^2}{1 + \alpha^2}
\label{eq:1s-chebyshev}\ee
where $\mu := \E X$ and $\sigma^2 := \E X^2 - \mu^2$, we find
\be
  \Pr\left(v_1(\lambda)^2 \ge \binom{k}{2} - \alpha \frac{k^2}{\sqrt{2}}\right)
  \ge \frac{\alpha^2}{1+\alpha^2}
\,.
\ee
Choosing (say) $\alpha=\frac{1}{2\sqrt{2}}$, we have $\Pr(v_1(\lambda)^2 \ge
\frac{1}{4}k^2 - \frac{1}{2}k) \ge \frac{1}{9}$; plugging this into
\eq{v1bound} gives 
\be
  \Delta_{k,d} \ge \frac{1}{18}
  \left[1-\exp\left(-\frac{k}{2d}\sqrt{1-\frac{2}{k}}\right)\right]
\,,
\ee
which is $\Omega(1)$ provided $k=O(d)$.  To obtain a specific numeric
bound, we divide into cases.  If $2\leq k \leq 5$, then $\Delta_{k,d}
\geq \Delta_{2,d} = 1/d \geq k/36d$, where the equality can be
verified by substituting $k=2$ into \eq{Delta-kd}.  If $6\leq k \leq d$, then
$\frac{1}{2}(1-\frac{2}{k})^{1/2} \geq \frac{1}{\sqrt{6}}$, so $\Delta_{k,d}
\geq\frac{1}{18}(1- e^{-k/\sqrt{6}d})$.  Next, the
convexity of $e^{-x/\sqrt{6}}$ 
implies that $e^{-x/\sqrt{6}}\geq 1-xe^{-1/\sqrt{6}} \geq 1-x/2$.
Thus $\Delta_{k,d} \geq k/36d$.
\end{proof}

To put the results of this section in context, we review the situation
when $k\gg d$ or $k \ll d$.  When $k\ra\infty$ with $d$ held constant, 
applying the measurement $\{\Pi_\lambda\}_{\lambda \partitionof k}$ to
$\rho^{\ot k}$ and outputting $\bar{\lambda}:=\lambda/k$ has long been known to be a valid estimator of the
spectrum of $\rho$ \cite{KW01}. 
Indeed, if $r_1\geq \ldots \geq r_d$ are the eigenvalues of $\rho$, then \cites{Hayashi:02b,CM04} proved that
\be
  \tr \Pi_\lambda \rho^k \leq (k+1)^{d(d-1)/2} \exp\left(-kD(\bar{\lambda}
\| r)\right)
\,,
\ee
where $D(p\|q):=\sum_i p_i\log(p_i/q_i)$ is the (classical) relative
entropy.  In fact, this coefficient of $k$ in the exponential is
optimal as $k\ra \infty$.
This inequality is usually only interesting when $k=\Omega(d^2)$, so our \thmref{nonadaptive} can be viewed as the first positive
result for spectrum estimation in the regime where $k=o(d^2)$.

What about the other limit, when $k\ll d$?  The only constructive
result previously known was the swap test, which yields
$\Delta_{2,d}=1/d$.  However,
\be
  F(\schur(2,d),\planch(2))=\frac{1}{2}\left(\sqrt{1-\frac{1}{d}}
 + \sqrt{1+\frac{1}{d}}\right) = 1-O\left(1/d^2\right)
\,,
\ee
so repeating swap tests will actually require $k=\Omega(d^2)$ copies
before achieving $\Omega(1)$ distinguishability.  Indeed,
$k=\Omega(d^2)$ copies are needed for $\Omega(1)$ distinguishability
even if we perform the optimal individual measurement on $O(1)$ copies
 at a time.  This follows from the following variant of \lemref{schur-planch}:
\begin{lemma}
\ba
  F\left(\schur(k,d), \planch(k)\right) &\geq 1 - \frac{k^3}{12 d^2}
\label{eq:PS-fid-standard}\\
  F\left(\schur(k,d), \planch(k)\right) &\geq 1 - C^{-\sqrt{k}}
- \frac{9}{4}\frac{k^2}{d^2}
\label{eq:PS-fid-large-dev}
\,,
\ea
where $C>1$ is a universal constant.
\end{lemma}
These bounds are not quite strong enough to match the $1-O(k^2/d^2)$ upper
bound on the fidelity from \lemref{schur-planch} (which we conjecture is tight), but they still prove
that achieving the performance of the optimal measurement requires
operations that are entangled across many copies.

\begin{proof}
We begin with a direct calculation of
\be
  f_{k,d} := F\left(\schur(k,d), \planch(k)\right)
  = \E_{\lambda\partitionof k} \sqrt{\prod_{(i,j)\in\lambda}
  \left(1 + \frac{j-i}{d}\right)}
\,.
\label{eq:sp-fid-formula}
\ee
Consider the contribution from a particular $\lambda$, and suppose for now that $\lambda=\lambda^T$.  Then $(i,j)\in\lambda$ if and only if $(j,i)\in\lambda$, so each factor of $1+(j-i)/d$ is accompanied by a corresponding factor of $1-(j-i)/d$ (unless $i=j$, in which case this term is $1$ and can be ignored).  Combining these terms, the entire product can be written as
\be
   \sqrt{\prod_{(i,j)\in\lambda} \left(1 + \frac{j-i}{d}\right)}
 = \left[\prod_{(i,j)\in\lambda} 
   \left(1 - \frac{(j-i)^2}{d^2}\right)\right]^{\frac{1}{4}}.
\label{eq:sym-quadratic-d}
\ee
Otherwise, suppose that $\lambda\neq \lambda^T$.  Recall that $\lambda$
and $\lambda^T$ have the same probability under the Plancherel
distribution.  Thus the terms corresponding to $\lambda$ and
$\lambda^T$ appear in \eq{sp-fid-formula} with the same
coefficient.  Next we use the identity
$x+y\geq 2\sqrt{xy}$ (for $x,y\geq 0$) to show
\be
     \sqrt{\prod_{(i,j)\in\lambda}   \left(1 + \frac{j-i}{d}\right)}
   + \sqrt{\prod_{(i,j)\in\lambda^T} \left(1 + \frac{j-i}{d}\right)}
  \geq 2\left[\prod_{(i,j)\in\lambda} 
  \left(1 - \frac{(j-i)^2}{d^2}\right)\right]^{\frac{1}{4}}
\,.
\label{eq:asym-quadratic-d}
\ee
Combining \eq{sym-quadratic-d} and \eq{asym-quadratic-d}, we can
lower bound \eq{sp-fid-formula} as
\be
  f_{k,d} \geq \E_{\lambda \partitionof k}
  \left[\prod_{(i,j)\in\lambda} 
  \left(1 - \frac{(j-i)^2}{d^2}\right)\right]^{\frac{1}{4}}
  \ge 1 - \frac{k^3}{12d^2}
\label{eq:PS-fid-quadratic}\,,
\ee
where the last inequality uses the fact that the product is minimized for the diagram $\lambda=(k)$.  This proves \eq{PS-fid-standard}.

The proof of \eq{PS-fid-large-dev} is mostly the same, but also uses
bounds on large deviations of $\lambda_1$ for Plancherel-distributed
$\lambda$.  In particular, equation (1.20) of \cite{BDJ99} implies
that under $\planch(k)$,
\be
  \Pr(\lambda_1\geq 3\sqrt{k}) \leq C^{-\sqrt{k}}
\ee
for a universal constant $C>1$ (in fact, $C =
\exp(6\cosh^{-1}(3/2)+2\sqrt{5})\approx 28189$).  Since $(i,j)\in\lambda$ implies that
$j-i \leq \lambda_1-1$, we can lower bound the middle expression in
\eq{PS-fid-quadratic} to give
\be
  f_{k,d} \geq \Pr(\lambda_1\leq 3\sqrt{k}) 
  \left(1-\frac{9k}{d^2}\right)^{\frac{k}{4}}
  \geq 1 - C^{-\sqrt{k}} - \frac{9}{4}\frac{k^2}{d^2}
\,,
\ee
which completes the proof.
\end{proof}

It is interesting to consider the classical analog of applying the
swap test to the collision problem.  If we compare $k/2$ sets of 2
objects each, then estimating the number of collisions up to a
constant multiple requires $k=\Omega(d)$.  This is also quadratically
worse than the optimal collective estimation procedure on $k$ objects,
but for apparently different reasons. 

\subsection{A black box for the quantum collision problem}
\label{sec:adaptive}

We now define a version of this problem that allows for adaptive
algorithms by introducing a quantum oracle.  The oracle implements the
isometry 
\be
  |i\> \mapsto |i\>|\psi_{f(i)}\>\,,
\label{eq:aqcp_isometry}
\ee
where $\mathcal{B}:=\{|\psi_1\>,\ldots,|\psi_d\>\}$ is an arbitrary (unknown) orthonormal basis of $\C^d$ and $f$ is either a one-to-one function or an $r$-to-one function.  The goal is to determine which is the case using as few queries as possible.  In fact, we assume that the isometry is extended to a unitary operator $R$ acting on $\C^d\otimes \C^d$ by
\be
  |i\>|y\> \mapsto |i\> \, U|y\oplus f(i)\>\,,
\label{eq:aqcp_oracle}
\ee
where $U:=\sum_i\ket{\psi_i}\bra{i}$ is the unitary matrix effecting a
transformation from the standard basis to $\mathcal{B}$. We also
assume we can perform its inverse $R^\dag$.

By considering the case where the basis $\mathcal{B}$ (or equivalently $U$) is known, it is clear that the quantum lower bound for the usual collision problem \cite{AS04} implies an $\Omega(\sqrt[3]{d/r})$ lower bound on the number of queries for the quantum collision problem as well.  We present an algorithm for this problem that requires only $O(\sqrt[3]{d/r} \log d/r)$ queries, nearly matching the lower bound.

The basic tool used in this algorithm is the following amplified
version of the swap test:

\begin{lemma}[Amplified swap test with small disturbance]
\label{lem:swaptest}
Suppose we are given a quantum state $\rho$ satisfying
\be
  \frac{1}{2} \left\| \rho - \proj{\psi}\right\|_1 \leq \eps
\ee
where
\be
  \ket{\psi} = \sum_i a_i \ket{\alpha_i}^{\ot m}\ket{\beta_i}^{\ot
  m}\ket{\gamma_i}
\ee
with $\<\gamma_i|\gamma_j\> = \delta_{i,j}$ and $\sum_i |a_i|^2=1$.
We further assume that for each $i$, $\theta_i := \scalar{\alpha_i}{\beta_i}$ is
equal to either 0 or 1.
Then there exists a unitary operation $U$, implementable with $O(m)$
gates, such that
\be
  \frac{1}{2} \left\| U\rho U^\dag - \proj{\psi'}\right\|_1 \leq
  \eps + 2^{1-m/2}
\ee
where 
\be
  \ket{\psi'} := \sum_i a_i (-1)^{\theta_i}
  \ket{\alpha_i}^{\ot m}\ket{\beta_i}^{\ot m}\ket{\gamma_i}
\,.
\ee
\end{lemma}

\begin{proof}
The procedure for implementing $U$ is as follows.  First perform the $m$-fold swap test without the final measurements.  Conditioned on any swap test showing that the states are orthogonal, record this answer in an ancilla qubit; otherwise, record that they are the same.  Apply a phase of $-1$ in the ``same'' case.  Finally, uncompute the ancilla qubit along with all the swap tests.

To analyze $U$, we initially assume we are acting on $\proj{\psi}$.  
Let $\Pi$ denote the projector that results from each swap test
returning the symmetric outcome, so that $U = 1 - 2\Pi$.
Since the swap test will always project $\ket{\alpha_i}\ket{\alpha_i}$
onto the symmetric space, we have
\be
  \Pi\ket{\alpha_i}^{\ot m}\ket{\beta_i}^{\ot m}\ket{\gamma_i}
   = \ket{\alpha_i}^{\ot m}\ket{\beta_i}^{\ot m}\ket{\gamma_i}
\ee
whenever $\theta_i=1$.  On the other hand, if $\theta_i=0$ then
\be
  \Pi\ket{\alpha_i}^{\ot m}\ket{\beta_i}^{\ot m}\ket{\gamma_i}=
  \frac{1}{\sqrt{2^m}}\ket{\eta_i}^{\ot m}\ket{\gamma_i}
\ee
where
$\ket{\eta_i}:=\frac{1}{\sqrt{2}}\left(\ket{\alpha_i}\ket{\beta_i}
+ \ket{\beta_i}\ket{\alpha_i}\right)$.
Combining these, we obtain
\ba
  U\ket{\psi} &=
  \sum_i a_i \left( (-1)^{\theta_i}
  \ket{\alpha_i}^{\ot m} \ket{\beta_i}^{\ot m} -
  \delta_{\theta_i,0}
  2^{1-\frac{m}{2}} \ket{\eta_i}^{\ot m}\right)\ket{\gamma_i}\\
  &= \ket{\psi'} - \sum_i a_i\delta_{\theta_i,0}
  2^{1-\frac{m}{2}} \ket{\eta_i}^{\ot m}\ket{\gamma_i}
\,.
\ea
Now the fidelity between $U |\psi\>$ and $|\psi'\>$ is
\ba
  F\left(|\psi'\>\<\psi'|,U|\psi\>\<\psi|U^\dag\right) & = 
  \left|\bra{\psi'}U\ket{\psi}\right|^2 =
  \left| 1 - \sum_i a_i \delta_{\theta_i,0} 2^{1-\frac{m}{2}}
  \<\psi'|\eta_i\>^{\ot m}\ket{\gamma_i}
  \right|^2 \\
  &=
  \left| 1 - \sum_i |a_i|^2 \delta_{\theta_i,0} 2^{1-m}\right|^2 
  \geq \left| 1-2^{1-m}\right|^2 \geq 1-2^{2-m}
\,.
\label{eq:amp-swap-fid-calc}
\ea
Finally,
\ba
  \frac{1}{2} \left\| U\rho U^\dag - \proj{\psi'} \right\|_1
  &\leq
  \frac{1}{2} \left\| U\rho U^\dag - U|\psi\>\<\psi|U^\dag\right\|_1 +
  \frac{1}{2} \left\| U|\psi\>\<\psi| U^\dag - \proj{\psi'}\right\|_1
\label{eq:amp-swap-tri-ineq}\\
  &\leq
  \frac{1}{2} \left\| \rho - \proj{\psi}\right\|_1 + 2^{1-\frac{m}{2}}
\label{eq:amp-swap-fid-trace}
   \leq \eps + 2^{1-\frac{m}{2}}
\,,
\ea
where \eq{amp-swap-tri-ineq} uses the triangle inequality and
\eq{amp-swap-fid-trace} uses first the unitary invariance of the trace
norm and then \eq{amp-swap-fid-calc} and \eq{fidelity-trace-dist}.
\end{proof}

We will use this amplified swap test to prove

\begin{theorem}
The quantum $r$-collision problem has query complexity $O(\sqrt[3]{d/r} \log d/r)$.
\end{theorem}
\begin{proof}
We give an explicit algorithm based on the quantum algorithm for the usual collision problem \cite{BHT97}.  We first outline that original algorithm and then show how to extend it to the quantum problem.

The algorithm builds a table of the first $\sqrt[3]{d/r}$ items and uses Grover's algorithm to search the remaining items for an entry of the table.  Set $K:=\{1,\ldots,\sqrt[3]{d/r}\}$ and $\bar{K}:=\{1,\ldots,d\}\setminus K$.  Then the algorithm consists of two steps:
\begin{enumerate}
\item Build the table $L:=\{(i,f(i)) : i\in K\}$.
\item Use Grover search to find $j\in\bar{K}$ such that $F(j)=1$, where the function $F : \bar{K} \rightarrow \{0,1\}$ is defined by $F(j)=1$ if there is $i\in K$ with $f(i)=f(j)$ and $F(j)=0$ otherwise.
\end{enumerate}
The first step uses $O(\sqrt[3]{d/r})$ queries.  In the second step, the function $F$ must be computed reversibly, so two queries ($R$ and $R^\dag$) are used for each call to $F$.  Note that it is important that $F$ can be computed on a superposition of inputs.  The total number of queries in the second step is twice the number of Grover iterations needed to find a marked item, which we now analyze.

Assume that $f$ is an $r$-to-one function.  The probability that there
is any collision among the elements of $K$ is only $O((r/d)^{1/6})$,
implying that the number of $j \in \bar K$ with $F(j)=1$ is
$(r-1) (d/r)^{1/3}$ with high probability.  Therefore, the number of Grover iterations necessary to find a solution is $O(\sqrt{d/[r(d/r)^{1/3}]})=O((d/r)^{1/3})$.  In
summary, the total number of queries of the above algorithm is
$O((d/r)^{1/3})$. 

Now we adapt this algorithm to the quantum collision problem.  Some
adaptation is necessary because it is not possible to directly check 
equality of two quantum states $|\psi_j\>$ and $|\psi_{j'}\>$.
However, using \lemref{swaptest}, we can effectively test equality
using $m:=2+2\log d/r$ copies of the quantum states, increasing the query
complexity only by a factor of $O(\log d/r)$. 

For this to work, it is important that we can reuse the states
corresponding to the entries in the table.  Iterating
\lemref{swaptest}, we find that the error after $\ell$
Grover iterations is at most $\ell \cdot 2^{1-m/2} \leq \ell r/d$.  Since the
number of Grover iterations is $\ell=O((d/r)^{1/3})$, the total error is
asymptotically negligible, and we obtain essentially the same
performance as in the classical collision problem.
\end{proof}

So far, we have only considered the query complexity of the quantum collision problem, without worrying about the algorithm's running time.  The quantum algorithm for the classical collision problem can run in time $O(\sqrt[3]{d/r} \log d/r)$ (i.e., only logarithmically greater than the query complexity), since after sorting the table $L$ (in time $O(\sqrt[3]{d/r} \log d/r)$), the function $F$ can be computed in time $O(\log d/r)$ \cite{BHT97}.  For the quantum collision problem, there is no obvious way to preprocess the table of quantum states.  However, by using Grover's algorithm to implement $F$ in time $O(\sqrt{(d/r)^{1/3}} \log d/r)$, we can obtain a running time of $O(\sqrt{d/r} \log d/r)$---substantially greater than the query complexity, but better than implementing $F$ by brute force search.  Note that to achieve these running times (for the both classical and quantum collision problems), we must assume a quantum RAM model.  In other words, we must assume that it is possible to access any desired location in the table---or any superposition thereof---in constant time (see for example \cite{NC00}*{Section 6.5}).  If we restrict ourselves to the quantum circuit model, not allowing parallelization, then the running time is even longer for both problems.

We conclude this section by briefly considering some alternate versions
of the quantum collision problem, one that has a fairly simple solution, and two about which less is known.
\begin{itemize}
\item {\em Input in an unknown basis:}
  Our definition of the black box for the
  quantum collision problem in \eq{aqcp_oracle} gives the output in an unknown
  basis.  Alternatively, we can consider a problem where the input is taken
  in an unknown basis.  Here we define the oracle $R$ to act as
  $R\left[U\ket{i}\ot \ket{y}\right] = U\ket{i} \ot \ket{y\oplus
  f(i)}$, where $U$ is again an unknown unitary and $f$ is either
  one-to-one or $r$-to-one, and the problem is to determine which is the case.
  This problem was suggested by Scott Aaronson;
  the solution we give here was also found independently by Andris Ambainis.

  The algorithm first queries $R$ on $(d/r)^{1/3}$ basis states of the form
  $\ket{j} \otimes \ket{0}$ for uniformly random $j$, and then measures the
  second register, producing a table $L$ of uniformly random function values
  $f(i)$, which are all distinct with high probability.  Then we 
  perform a Grover search for a state $U\ket{i}$ such that $f(i)\in L$.
  There are two components to the Grover search.  First, we need to
  identify our target space.  Let $\Pi:=\sum_{i\in
  f^{-1}(L)}U\proj{i}U^\dag$.  Note that $\tr \Pi$ is either $(d/r)^{1/3}$
  or $r(d/r)^{1/3}$, and that we can perform the reflection
  $1-2\Pi$ using only two queries (one of $R$ and one of $R^\dagger$).
  This implementation can be done by brute force in time $O(r(d/r)^{1/3})$, 
  or by Grover search in time $O(\sqrt{r(d/r)^{1/3}})$---either way, regardless  
  of whether $f$ is one-to-one or $r$-to-one, and using only two queries.
  Next, we need a starting state $\ket{\psi}$
  that we can efficiently map to and from the $\ket{0}$ state, so
  that we can both initialize it and perform the reflection
  $1-2\proj{\psi}$.  One such $\ket{\psi}$ is a state
  that is maximally entangled between the first register and an
  ancilla register of equal dimension.  Then we have $\bra{\psi}(I\ot
  \Pi)\ket{\psi} = \tr\Pi / d$.

  The algorithm now alternates $\ell$ times between the operations
  $1-2\proj{\psi}$ and $1-2\Pi$.  We choose $\ell$ so that if
  $f$ is $r$-to-one then the probability of finding a solution is
  nearly $1$, while if $f$ is one-to-one then the probability is far
  from one.  Thus, we will be able to distinguish the two cases with
  constant advantage.  Following \cite{BHMT02}, we
  set 
  \be
    \ell:=\left\lfloor 4/{\pi\sqrt{\sin^{-1}\left(\left({r}/{d}\right)^{\frac{2}{3}}\right)}}\right\rfloor
    = \Theta((d/r)^{1/3})
  \,.
  \ee
  The probability of finding a solution is $\geq 1 - O(r^{1/3}/d^{2/3})$ if $f$
  is $r$-to-one and is $\leq 1/r + O(r^{1/3}/d^{2/3})$ if $f$ is
  one-to-one. 

  If instead both the input and output bases are unknown, then we know
  of no algorithm better than making $O(d)$ queries with random inputs
  and using the algorithm of \secref{nonadaptive}.

\item {\em No access to $R^\dag$:}
  Suppose that instead of the full
  unitary oracle \eq{aqcp_oracle}, $R$ is simply the isometry
  \eq{aqcp_isometry}, mapping $\ket{i}$ to $\ket{i}\ket{\psi_{f(i)}}$.  
  In this case, the best algorithm we know queries the oracle 
  $O(\log d)$ times on each of $\sqrt{d}$ inputs (so that there is a 
  collision among those inputs with reasonably high probability), 
  and then searches the resulting table (either by brute force or by Grover
  search; in either case, making no further queries) using \lemref{swaptest}
  to perform comparisons.
  We conjecture that this is optimal (up to the $\log$ factor), but we 
  know of no lower bound better than the $\Omega(d^{1/3})$ bound from
  the classical problem.  

\item {\em Black box rank estimation:}
  Suppose again that $R$ is an isometry, but now mapping $\ket{i}$ to
  $\ket{i}\ket{\psi_i}$, and we are only promised that either
  $\{\ket{\psi_i}\}$ forms a basis for $\bbC^d$ or that
  $\frac{1}{d}\sum_i \proj{\psi_i}$ is proportional to a projector onto
  a $(d/r)$-dimensional subspace.  Here the best algorithm we know is to
  query $R$ on $O(d)$ inputs and use the rank estimation
  procedure from \secref{nonadaptive}.  We conjecture that this is
  optimal, but again have only the lower bound of $\Omega(d^{1/3})$.
\end{itemize}

\section{Discussion}

In this article, we have shown that weak Fourier-Schur sampling typically provides insufficient information to solve the hidden subgroup problem.  Nevertheless, it remains possible that Schur duality could be a useful tool for the \HSP.  Just as weak Fourier sampling can always be used to refine the state space into smaller subspaces, even when it alone fails to solve the \HSP, so we can use weak Fourier-Schur sampling to decompose the space even further.  The Schur decomposition has the additional complication that the refined subspaces are no longer simply tensor products of single-copy subspaces, but this may actually be an advantage since entangled measurements are known to be necessary for some groups.

In principle, strong Fourier-Schur sampling is guaranteed to provide enough information to solve the \HSP, simply because the hidden subgroup states are always distinguishable with $k=\poly(\log |G|)$ copies.  However, it would be  interesting to find a new efficient quantum algorithm for some \HSP based on strong Fourier-Schur sampling.  Perhaps a first step in this direction would be to analyze the performance of measurement in a random basis, as has been studied extensively in the case of weak Fourier sampling \cites{GSVV04,MRRS04,RRS05,Sen05}.

Moving away from our original motivation of the hidden subgroup problem, the quantum collision problem may be of independent interest.  As discussed in the latter part of \secref{nonadaptive}, our results on the quantum collision sampling problem can be viewed as an exploration of spectrum estimation with $k = o(d^2)$ copies, but much remains unknown about that regime.  Many open problems also remain regarding the black box version of the quantum collision problem: Can the running time be reduced?  What if both input and output registers are in unknown bases?  Is the inverse oracle essential?  And is black box rank estimation really no easier than rank estimation by sampling?

\section*{Acknowledgments}

We thank Scott Aaronson, Andris Ambainis, Masahito Hayashi, Keiji Matsumoto, Pranab Sen, and Umesh Vazirani for helpful discussions.
We also thank Patrick Hayden for organizing a Bellairs Research Institute workshop on representation theory in quantum information, at which the seeds for this work were planted.
This work was supported in part by the National Science Foundation
under grant PHY-456720, by the Army Research Office under grant
W9111NF-05-1-0294, and by the European Commission under Marie Curie
grant ASTQIT.



\begin{bibdiv}
\begin{biblist}

\bib{AS04}{article}{
      author={Aaronson, S.},
      author={Shi, Y.},
       title={Quantum lower bounds for the collision and the element
              distinctness problems},
     journal={J. ACM},
        year={2004},
      volume={51},
      number={4},
       pages={595\ndash 605},
        note={\eprint{quant-ph/0111102}, \eprint{quant-ph/0112086}},
}

\bib{BCD05}{inproceedings}{
      author={Bacon, D.},
      author={Childs, A. M.},
      author={van Dam, W.},
       title={From optimal measurement to efficient quantum algorithms
              for the hidden subgroup problem over semidirect product
              groups},
   booktitle={Proc. 46th IEEE Symposium on
              Foundations of Computer Science},
       pages={469\ndash 478},
        year={2005},
      eprint={quant-ph/0504083},
}

\bib{BCD05b}{article}{
      author={Bacon, D.},
      author={Childs, A. M.},
      author={van Dam, W.},
       title={Optimal measurements for the dihedral hidden subgroup problem},
     journal={to appear in Chicago Journal of Theoretical Computer Science},
      eprint={quant-ph/0501044},
}

\bib{BCH04}{techreport}{
      author={Bacon, D.},
      author={Chuang, I. L.},
      author={Harrow, A. W.},
       title={Efficient quantum circuits for {S}chur and {C}lebsch-{G}ordan
              transforms},
      eprint={quant-ph/0407082},
}

\bib{BCH06}{inproceedings}{
      author={Bacon, D.},
      author={Chuang, I. L.},
      author={Harrow, A. W.},
       title={The quantum {S}chur transform: {I}. {E}fficient qudit circuits},
   booktitle={to appear in Proc. 18th ACM-SIAM Symposium on Discrete Algorithms},
        year={2007},
      eprint={quant-ph/0601001},
}

\bib{BBMR05}{article}{
author = {Bagan, E.},
author = {Ballester, M. A},
author = {Munoz-Tapia, R.},
author = {Romero-Isart, O.},
title = {Purity estimation with separable measurements},
eprint = {quant-ph/0509087},
journal = {Phys. Rev. Lett.},
volume = {95},
pages = {110504},
year = {2005},
}

\bib{BDJ99}{article}{
      author={Baik, J.},
      author={Deift, P.},
      author={Johansson, K.},
       title={On the distribution of the length of the longest increasing
              subsequence of random permutations},
     journal={J. AMS},
      volume={12},
      number={4},
       pages={1119\ndash 1178},
        year={1999},
      eprint={math.CO/9810105},
}

\bib{BBDEJM96}{article}{
   author={A. Barenco},
   author={A. Berthiaume},
   author={D. Deutsch},
   author={A. Ekert},
   author={R. Jozsa},
   author={C. Macchiavello},
    title={Stabilisation of quantum computations by symmetrisation},
  journal={SIAM J. Comput.},
    pages={1541\ndash 1557},
     year={1997},
   eprint={quant-ph/9604028},
}

\bib{Bea97}{inproceedings}{
      author={Beals, R.},
       title={Quantum computation of {F}ourier transforms over symmetric
              groups},
        date={1997},
   booktitle={Proc. 29th ACM Symposium on Theory of Computing},
   publisher={ACM Press},
     address={New York},
       pages={48\ndash 53},
}

\bib{BL95}{inproceedings}{
      author={Boneh, R.},
      author={Lipton, R.},
       title={Quantum cryptanalysis of hidden linear functions},
        date={1995},
   booktitle={Proc. Advances in Cryptology,
              Lecture Notes in Computer Science {\bf 963}},
       pages={424\ndash 437},
}

\bib{BHT97}{inproceedings}{
      author={Brassard, G.},
      author={H{\o}yer, P.},
      author={Tapp, A.},
       title={Quantum cryptanalysis of hash and claw-free functions},
      eprint={quant-ph/9705002},
   booktitle={Proc. 3rd Latin American Symposium on Theoretical Informatics,
              Lecture Notes in Computer Science {\bf 1380}},
       pages={163\ndash 169},
        year={1998},
}

\bib{BHMT02}{incollection}{
author = {G. Brassard},
author = {H{\o}yer, P.},
author = {Mosca, M.},
author = {Tapp, A.},
title = {Quantum amplitude amplification and estimation},
eprint = {quant-ph/0005055},
booktitle = {Quantum Computation \& Information},
publisher = {AMS},
series = {Contemporary Mathematics Series Millenium Volume},
editor = {S. J. Lomonaco},
editor = {H. E. Brandt},
volume = {305},
pages = {53\ndash 74},
year = {2002},
}

\bib{BCWW01}{article}{
      author={H. Buhrman},
      author={R. Cleve},
      author={J. Watrous},
      author={de Wolf, R.},
       title={Quantum fingerprinting},
     journal={Phys. Rev. Lett.},
      volume={87},
       pages={167902},
        year={2001},
      eprint={quant-ph/0102001},
}

\bib{CW05}{techreport}{
      author={Childs, A. M.},
      author={Wocjan, P.},
       title={On the quantum hardness of solving isomorphism problems as
              nonabelian hidden shift problems},
      eprint={quant-ph/0510185},
}


\bib{Cop94}{techreport}{
      author={D. Coppersmith},
       title={An approximate {F}ourier transform useful in quantum
              factoring},
 institution={IBM Research Division},
     address={Yorktown Heights, NY},
      number={RC 19642},
        year={1994},
      eprint={quant-ph/0201067},
}

\bib{CM04}{article}{
  author = {Christandl, M.},
  author = {Mitchison, G.},
  title = {The spectra of density operators and the {K}ronecker coefficients of the symmetric group},
eprint = {quant-ph/0409016},
  year = {2006},
journal = {Commun. Math. Phys.},
volume = {261},
number = {3},
pages ={789\ndash 797},
}

\bib{EH99}{techreport}{
      author={Ettinger, M.},
      author={H{\o}yer, P.},
       title={A quantum observable for the graph isomorphism problem},
      eprint={quant-ph/9901029},
}

\bib{EH00}{article}{
      author={Ettinger, M.},
      author={H{\o}yer, P.},
       title={On quantum algorithms for noncommutative hidden subgroups},
        date={2000},
     journal={Advances in Applied Mathematics},
      volume={25},
      number={3},
       pages={239\ndash 251},
      eprint={quant-ph/9807029},
}

\bib{EHK99}{techreport}{
      author={Ettinger, M.},
      author={H{\o}yer, P.},
      author={Knill, E.},
       title={Hidden subgroup states are almost orthogonal},
      eprint={quant-ph/9901034},
}

\bib{FIMSS03}{inproceedings}{
      author={Friedl, K.},
      author={Ivanyos, G.},
      author={Magniez, F.},
      author={Santha, M.},
      author={Sen, P.},
       title={Hidden translation and orbit coset in quantum
              computing},
        date={2003},
   booktitle={Proc. 35th ACM Symposium on Theory of Computing},
   publisher={ACM Press},
     address={New York},
       pages={1\ndash 9},
      eprint={quant-ph/0211091},
}

\bib{FG97}{article}{
    author = {C.~A.~Fuchs},
    author = {J.~van~de~Graaf},
     title = {Cryptographic distinguishability measures for quantum
              mechanical states},
    eprint = {quant-ph/9712042},
   journal = {IEEE Trans. Inf. Theory},
    volume = {45},
    number = {4},
     pages = {1216\ndash 1227},
      year = {1999},
}

\bib{Gav04}{article}{
      author={Gavinsky, D.},
       title={Quantum solution to the hidden subgroup problem for
              poly-near-{H}amiltonian groups},
        date={2004},
     journal={Quantum Information and Computation},
      volume={4},
      number={3},
       pages={229\ndash 235},
}

\bib{GW98}{book}{
      author={Goodman, R.},
      author={Wallach, N. R.},
       title={Representations and Invariants of the Classical Groups},
   publisher={Cambridge University Press},
     address={Cambridge},
        year={1998},
}

\bib{GSVV04}{article}{
      author={Grigni, M.},
      author={Schulman, L.},
      author={Vazirani, M.},
      author={Vazirani, U.},
       title={Quantum mechanical algorithms for the nonabelian hidden
              subgroup problem},
        date={2004},
     journal={Combinatorica},
      volume={24},
      number={1},
       pages={137\ndash 154},
}

\bib{Gro96}{inproceedings}{
      author={Grover, L. K.},
       title={A fast quantum mechanical algorithm for database search},
   booktitle={Proc. 28th ACM Symposium on Theory of Computing},
        year={1996},
       pages={212\ndash 219},
      eprint={quant-ph/9605043},
}

\bib{Hal02}{inproceedings}{
      author={Hallgren, S.},
       title={Polynomial-time quantum algorithms for {P}ell's equation and the
              principal ideal problem},
   booktitle={Proc. 34th ACM Symposium on Theory of Computing},
        year={2002},
       pages={653\ndash 658},
}

\bib{Hal05}{inproceedings}{
      author={Hallgren, S.},
       title={Fast quantum algorithms for computing the unit group and class
              group of a number field},
   booktitle={Proc. 37th ACM Symposium on Theory of Computing},
        year={2005},
       pages={468\ndash 474},
}

\bib{HH00}{inproceedings}{
      author={Hales, L.},
      author={Hallgren, S.},
       title={An improved quantum {F}ourier transform algorithm and
  applications},
        date={2000},
   booktitle={Proc. 41st IEEE Symposium on Foundations of Computer
              Science},
       pages={515\ndash 525},
}

\bib{HMRRS06}{inproceedings}{
      author={Hallgren, S.},
      author={Moore, C.},
      author={R{\"o}tteler, M.},
      author={Russell, A.},
      author={Sen, P.},
       title={Limitations of quantum coset states for graph isomorphism},
        date={2006},
   booktitle={Proc. 38th ACM Symposium on Theory of Computing},
       pages={604\ndash 617},
        note={\eprint{quant-ph/0511148}, \eprint{quant-ph/0511149}},
}

\bib{HRT00}{inproceedings}{
      author={Hallgren, S.},
      author={Russell, A.},
      author={Ta-Shma, A.},
       title={The hidden subgroup problem and quantum computation using
              group representations},
   booktitle={Proc. 32nd ACM Symposium on Theory of Computing},
        year={2000},
       pages={627\ndash 635},
}

\bib{Har05}{thesis}{
        author={A. W. Harrow},
         title={Applications of coherent classical communication and the
                {S}chur transform to quantum information theory},
          type={Ph.D. thesis},
   institution={Massachusetts Institute of Technology},
          year={2005},
        eprint={quant-ph/0512255},
}


\bib{Hayashi:02b}{article}{
    author={M. Hayashi and K. Matsumoto},
    title={Quantum universal variable-length source coding},
    journal= {Phys. Rev. A},
    volume= {66},
    year= {2002},
    pages={022311},
    number= {2},
eprint = {quant-ph/0202001}
}

\bib{IMS03}{article}{
      author={Ivanyos, G.},
      author={Magniez, F.},
      author={Santha, M.},
       title={Efficient quantum algorithms for some instances of the
              non-abelian hidden subgroup problem},
        date={2003},
     journal={International Journal of Foundations of Computer
              Science},
      volume={14},
      number={5},
       pages={723\ndash 739},
      eprint={quant-ph/0102014},
}

\bib{Juc74}{article}{
      author={Jucys, A.},
       title={Symmetric polynomials and the center of the symmetric group ring},
     journal={Rep. Math. Phys.},
      volume={5},
      number={1},
        year={1974},
       pages={107\ndash 112},
}

\bib{Ker93}{article}{
      author={Kerov, S.},
       title={Gaussian limit for the {P}lancherel measure of the symmetric
              group},
     journal={Comptes Rendus Acad. Sci. Paris, S{\'e}r. I},
      volume={316},
       pages={303\ndash 308},
        year={1993},
}

\bib{KW01}{article}{
      author={Keyl, M.},
      author={Werner, R. F.},
       title={Estimating the spectrum of a density operator},
     journal={Phys. Rev. A},
      volume={64},
       pages={052311},
        year={2001},
      eprint={quant-ph/0102027},
}

\bib{Kup03}{article}{
      author={Kuperberg, G.},
       title={A subexponential-time quantum algorithm for the dihedral
              hidden subgroup problem},
     journal={SIAM Journal on Computing},
      volume={35},
      number={1},
       pages={170\ndash 188},
        year={2005},
      eprint={quant-ph/0302112},
}

\bib{LS77}{article}{
      author={Logan, B. F.},
      author={Shepp, L. A.},
       title={A variational problem for random {Y}oung tableaux},
     journal={Adv. Math.},
      volume={26},
        year={1977},
      number={2},
       pages={206\ndash 222},
}

\bib{MRR04}{inproceedings}{
      author={Moore, C.},
      author={Rockmore, D.~N.},
      author={Russell, A.},
       title={Generic quantum {F}ourier transforms},
   booktitle={Proc. 15th ACM-SIAM Symposium on Discrete Algorithms},
        year={2004},
       pages={778\ndash 787},
      eprint={quant-ph/0304064},
}

\bib{MRRS04}{inproceedings}{
      author={Moore, C.},
      author={Rockmore, D.~N.},
      author={Russell, A.},
      author={Schulman, L.~J.},
       title={The hidden subgroup problem in affine groups: Basis
              selection in {F}ourier sampling},
        date={2004},
   booktitle={Proc. 15th ACM-SIAM Symposium on
              Discrete Algorithms},
   publisher={SIAM},
     address={Philadelphia},
       pages={1113\ndash 1122},
        note={\eprint{quant-ph/0211124},
              extended version available at
              \eprint{quant-ph/0503095}},
}

\bib{MRS05}{inproceedings}{
      author={Moore, C.},
      author={Russell, A.},
      author={Schulman, L.~J.},
       title={The symmetric group defies strong {F}ourier sampling},
   booktitle={Proc. 46th IEEE Symposium on
              Foundations of Computer Science},
        year={2005},
       pages={479\ndash 490},
      eprint={quant-ph/0501056},
}

\bib{NC00}{book}{
      author={Nielsen, M. A.},
      author={Chuang, I. L.},
       title={Quantum Computation and Quantum Information},
   publisher={Cambridge University Press},
     address={Cambridge},
        year={2000}
}

\bib{RRS05}{inproceedings}{
      author={Radhakrishnan, J.},
      author={R{\"o}tteler, M.},
      author={Sen, P.},
       title={On the power of random bases in {F}ourier sampling: 
              Hidden subgroup problem in the {H}eisenberg group},
   booktitle={Proc. International Colloquium on Automata, Languages and
              Programming, Lecture Notes in Computer Science {\bf 3580}},
       pages={1399\ndash 1411},
        year={2005},
      eprint={quant-ph/0503114},
}

\bib{Reg02}{inproceedings}{
      author={Regev, O.},
       title={Quantum computation and lattice problems},
        date={2002},
   booktitle={Proc. 43rd Symposium on Foundations
              of Computer Science},
   publisher={IEEE},
     address={Los Alamitos, CA},
       pages={520\ndash 529},
      eprint={cs.DS/0304005},
}

\bib{Reg04}{techreport}{
      author={Regev, O.},
       title={A subexponential time algorithm for the dihedral hidden
              subgroup problem with polynomial space},
      eprint={quant-ph/0406151},
}

\bib{SV05}{inproceedings}{
      author={Schmidt, A.},
      author={Vollmer, U.},
       title={Polynomial time quantum algorithm for the computation of the unit
              group of a number field},
   booktitle={Proc. 37th ACM Symposium on the Theory of Computing},
        year={2005},
       pages={475\ndash 480},
}

\bib{Sel40}{article}{
      author={Selberg, H.~L.},
       title={Uber eine {U}ngleichungen zur {E}rg{\"a}nzung des
  {T}chebycheffschen {L}emmas},
        date={1940},
     journal={Skandinavisk Aktuarietidskrift},
      volume={23},
       pages={121\ndash 125},
}

\bib{Sen05}{techreport}{
      author={Sen, P.},
       title={Random measurement bases, quantum state distinction and
              applications to the hidden subgroup problem},
      eprint={quant-ph/0512085},
}

\bib{Ser77}{book}{
      author={Serre, J. P.},
       title={Linear Representations of Finite Groups},
   publisher={Springer},
     address={New York},
      series={Graduate Texts in Mathematics},
      volume={42},
        year={1977},
}

\bib{Sho97}{article}{
      author={Shor, P.~W.},
       title={Polynomial-time algorithms for prime factorization and
              discrete logarithms on a quantum computer},
        date={1997},
     journal={SIAM Journal on Computing},
      volume={26},
      number={5},
       pages={1484\ndash 1509},
      eprint={quant-ph/9508027},
}

\bib{Sta71}{article}{
      author={Stanley, R. P.},
       title={Theory and application of plane partitions},
     journal={Studies in Appl. Math.},
      volume={1},
       pages={167\ndash 187 and 259\ndash 279},
        year={1971},
}

\bib{VK77}{article}{
      author={Vershik, A.~M.},
      author={Kerov, S.~V.},
       title={Asymptotic behavior of the {P}lancherel measure of the symmetric
              group and the limit form of Young tableaux},
     journal={Dokl. Akad. Nauk SSSR},
      volume={233},
      number={6},
        year={1977},
       pages={1024\ndash 1027},
}

\bib{VK89}{article}{
      author={Vershik, A.~M.},
      author={Kerov, S.~V.},
       title={Asymptotic behavior of the maximum and generic
              dimensions of irreducible represenations of the
              symmetric group},
     journal={Funct. Anal. Appl.},
      volume={19},
       pages={21\ndash 31},
        year={1989},
        note={English translation of Funk. Anal. i Prolizhen {\bf 19}
              (1985), no. 1, 25\ndash 26},
}

\end{biblist}
\end{bibdiv}
\end{document}